\documentclass[%
 reprint,
 amsmath,amssymb
 aps,
]{revtex4-2}
\usepackage{color}

\usepackage{graphicx}
\usepackage{dcolumn}
\usepackage{bm}
\usepackage{multirow}

\begin{document}

\preprint{APS/123-QED}

\title{Supremacy of optimal beam energy for synthesis of superheavy elements}
\date{\today}
\author{H.C. Manjunatha$^{1}$, N. Sowmya$^{1}$, P.S.Damodara Gupta$^{1}$, L. Seenappa$^1$ and T. Nandi$^{2\dagger}$}
\affiliation {\it $^1$Department of Physics, Government College for women, Kolar-563101, Karnataka, India}
\affiliation {\it $^2$ 1003 Regal, Mapsko Royal Ville, Sector-82, Gurgaon, Haryana-122004, India}
\thanks{Formerly with: Inter-University Accelerator Centre, Aruna Asaf Ali Marg, New Delhi, 110067, India.\\$\dagger$ For correspondences: nanditapan@gmail.com}
\begin{abstract}
Besides right choice of entrance channel, selection of optimal beam energies for synthesis of superheavy elements plays a crucial role. A thorough investigation with the advanced statistical and dinuclear system models on all the experiments performed for the synthesis of the successful superheavy elements $Z=104-118$ and failed superheavy elements $Z=119-120$ leads us to infer that improper choice of the beam energies may be responsible for too low production cross sections to measure and thus the cause for the debacle. We have predicted the optimal beam energies to obtain the maximum production cross sections for all the reactions used for the superheavy elements $Z=104-120$. Hope exploitation of these  predictions may be on the cards soon to extend the periodic table for the eighth period. 
\end{abstract}
\maketitle
Superheavy elements (SHE) up to oganesson ($^{294}_{118}$Og) have been discovered using the cold and hot fusion reactions \cite{haba2019new}, which has completed up to the seventh row of the periodic table \cite{nazarewicz2018limits}. Currently, the synthesis of the SHE for Z $>$118 has posed a higher challenge for nuclear physics research as initial experimental attempts to discovering the elements $Z = 119$ and 120, the first two elements of the eighth row, have remained unsuccessful at the laboratories FLNR Dubna and GSI Darmstadt \cite{dullmann2016search,oganessian2009attempt,hofmann2015super}. The failure contradicts to the fact that the increasing stability is expected for the SHE Z$>$110 \cite{hamilton2013search,oganessian2015super}. Therefore, it may be inferred that the evaporation residue cross-sections ($\sigma_{ER}$) can be too low to measure. Hence, special effort is being invested to revamp the detection sensitivity close to fb. Whatsoever, it is vital to understand the reaction dynamics in order to choose the best reaction as well as experimental condition to produce a new SHE. Accordingly, we have taken up a theoretical project to address this important issue.\\
\indent Last year, extensive measurements of mass-angle distributions and cross sections for the cold fusion reactions $^{48}$Ca, $^{50}$Ti, and $^{54}$Cr + $^{208}$Pb \cite{banerjee2019mechanisms} have shown a systematic decrease in the fraction of mass-symmetric fission as well as $\sigma_{ER}$ as a function of increasing projectile charge and beam energy. One of our recent studies reveals the fact of the charge asymmetry with the hot fusion reactions also \cite{manjunatha2020entrance}. This year, another experiment \cite{tanaka2020study} on the excitation functions of quasielastic scattering cross sections for some hot fusion reactions. $^{22}Ne + ^{248}Cm, ^{26}Mg + ^{248}Cm$, and $^{48}Ca + ^{238}U$ used for synthesizing the SHE demonstrates the importance of the deformation of the target nuclei. Further, it shows that the optimum incident energy to have the highest production cross section is important for the hot fusion reactions and it can be determined by an experimentally observed barrier distribution. Note that practically, we can fabricate the elemental nuclear target up to californium ($Z=98$), therefore choice of the projectile with nuclear charge lower than $Z=21$ for synthesizing the SHE $Z>118$ is impossible. If we choose the lightest projectile then the target must be around the californium only; hence no choice is left with the deformation free targets. Whereas we have full freedom to choose the optimum beam energy for obtaining the largest $\sigma_{ER}$ for synthesizing a SHE of our choice. Furthermore, heavy-ion fusion reactions show unexpected behavior at sub-barrier energies, for example, a large increase in the $\sigma_{ER}$ is observed \cite{jiang2002unexpected}, even though the survival probability of superheavy nuclei (SHN) is a result of strong binding shell effect against the large Coulomb repulsion. The increased excitation energy reduces the shell effect which in turn decreases the survival probability of compound nucleus \cite{feng2007formation}. Therefore, the optimal beam energy may be only about 20–30 MeV higher than the corresponding Coulomb barrier for the production of neutron rich superheavy nuclides \cite{zagrebaev2011production}. However, the optimal beam energy for the production of neutron-rich isotopes of SHE in multi-nucleon transfer reactions with heavy actinide nuclei (like U+Cm) is very close to Coulomb barrier \cite{zagrebaev2013future}. Sometimes the optimal beam energy may even be at the sub-barrier energies \cite{dvorak2008observation,itkis2015fusion}.\\
\indent The synthesis of the superheavy elements is adjudged by the figure of the $\sigma_{ER}$; hence precise theoretical prediction of $\sigma_{ER}$ (the order of one picobarn) is crucial for conducting the experiments for synthesis of the SHE \cite{hofmann2015super}. A small increase of the $\sigma_{ER}$ could save months of expensive accelerator time. To this extent, the first question one might ask is, how exactly can we predict $\sigma_{ER}$ with current models? At present, the theoretical models such as advanced statistical model (ASM) \cite{d1994sensitivity,sagaidak1998fission} and dinuclear system (DNS) model \cite{adamian1996effective,fazio2004formation} are used for the excitation function studies. 
In this letter, firstly we have examined the ASM and DNS model predictions through both the cold and hot fusion experiments employed for the synthesis of SHN with $Z=104-118$. Next we have theoretically studied the excitation functions for all these reactions using both the ASM and DNS models, which gives us idea on the optimal energies to have the highest production cross section for a certain reaction. Finally we find the main reason for the failure of the synthesis of SHN $Z=119$ and 120. We believe exploitation of the present work will help greatly in synthesizing the SHE of the eighth period.\\
\begin{figure}
    \centering
    \includegraphics[width=\linewidth]{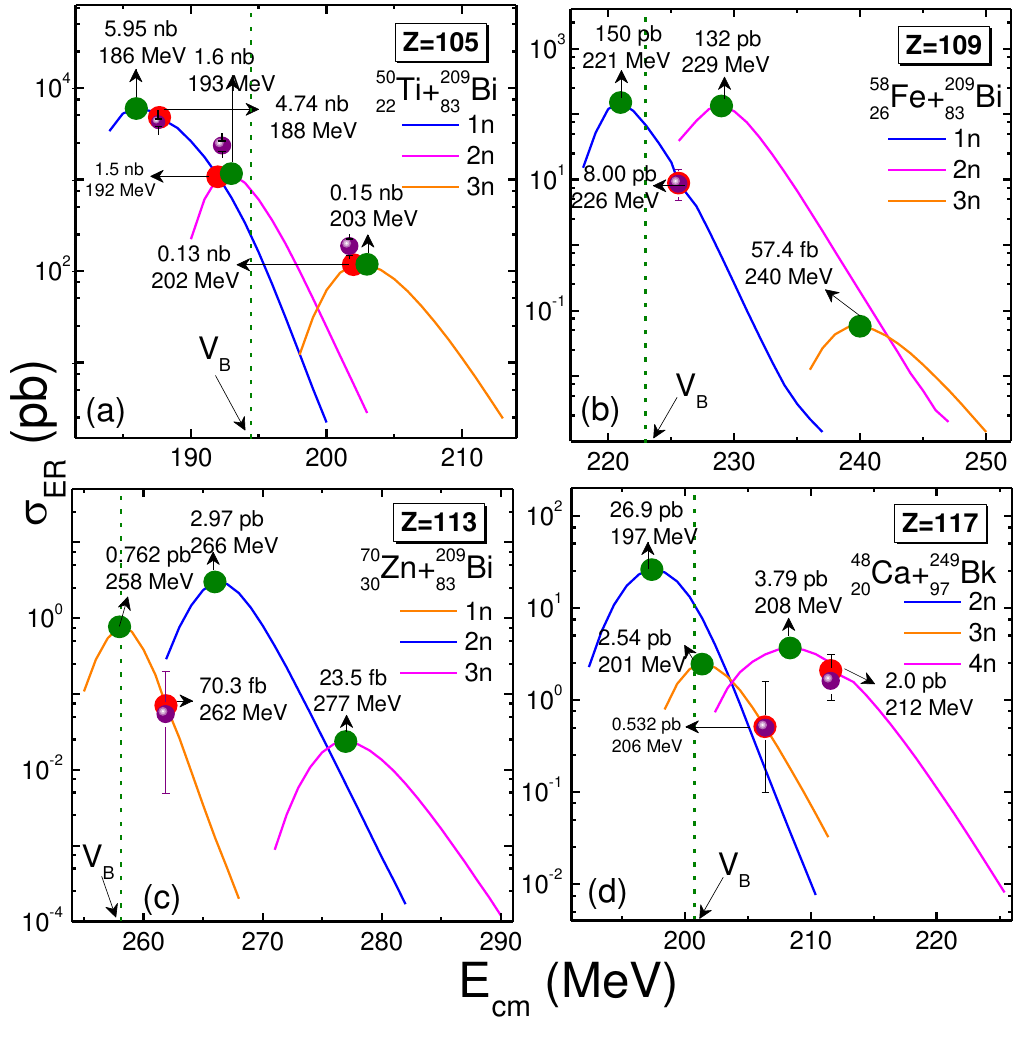}
    \caption{Excitation function plot for four fusion reactions used in the synthesis of superheavy elements $Z= 105, 109, 113$ and 117, where $\sigma_{ER}$ are calculated using the ASM.  Green  circles indicate the maximum cross-section predicted for a certain neutron evaporation channel. Violet circles represent the $\sigma_{ER}$ measured at energies given along side. Red circles specify the ASM predicted $\sigma_{ER}$ at the beam energy used for the experiments. Vertical green dotted line represents the fusion barrier as predicted by Ref. \cite{nandi2020search}.}
    \label{Exc-func-ASM}
\end{figure}
\indent We set to examine here the supremacy of optimal energy for synthesis of superheavy elements with the help of   the ASM and DNS models. According to the ASM, the evaporation residue cross section ($\sigma_{ER}$) for SHN production in a heavy-ion fusion reaction with subsequent emission of $x$ neutrons is given by \cite{sridhar2018search,manjunatha2018investigations,sridhar2019studies}
\begin{align}
    \sigma_{ER}^{xn}=\frac{\pi}{k^2}\sum_{\ell=0}^\infty(2\ell+1)T(E,\ell)P_{CN}(E_{cm},\ell)P_{sur}^{xn}(E^*,\ell) \label{sigmaEr-ASM}.
    \end{align}
\noindent Where $k$ is the wave number, average angular momentum $\ell=\langle\ell\rangle$, $P_{sur}$ is the survival probability of the compound nucleus in the ground state by emitting neutrons or lighter particles, $P_{CN}$ is the compound nucleus formation probability, the excitation energy $E^*=E_{cm}+Q$, the reaction $Q$-value is calculated \cite{wang2017ame2016} and $T_\ell$ is the $E_{cm}$ and $\ell$ dependent barrier penetration factor \cite{sridhar2018search,manjunatha2018investigations,sridhar2019studies}. The value of $x$ depends on the $E^*$ used for the reaction. While the $\sigma_{ER}$ by the DNS model \cite{giardina2000effect} is stated as 
\begin{align}
    \sigma_{ER}^{Z,A}=\sum_{\ell=0}^{\infty} (2\ell+1)\sigma_{cap}(E_{cm})W_{sur}^{Z,A}(E_{cm},J) \label{sigmaEr-DNS}.
    \end{align}
Where $\sigma_{cap}$ is the partial capture cross section which represents the transition of the colliding nuclei over the
Coulomb barrier and the formation of the initial DNS when the kinetic energy $E_{cm}$ and $\ell$
of the relative motion are transformed into the excitation energy and $\ell$ of the DNS. The probability
of the production of certain residual nucleus (Z,A) from the excited entrance channel (DNS) into a distinct decay
channel is described by the survival probability $W_{sur}^{Z,A}(E_{cm}, J)$.\\
\begin{figure}
    \centering
    \includegraphics[width=\linewidth]{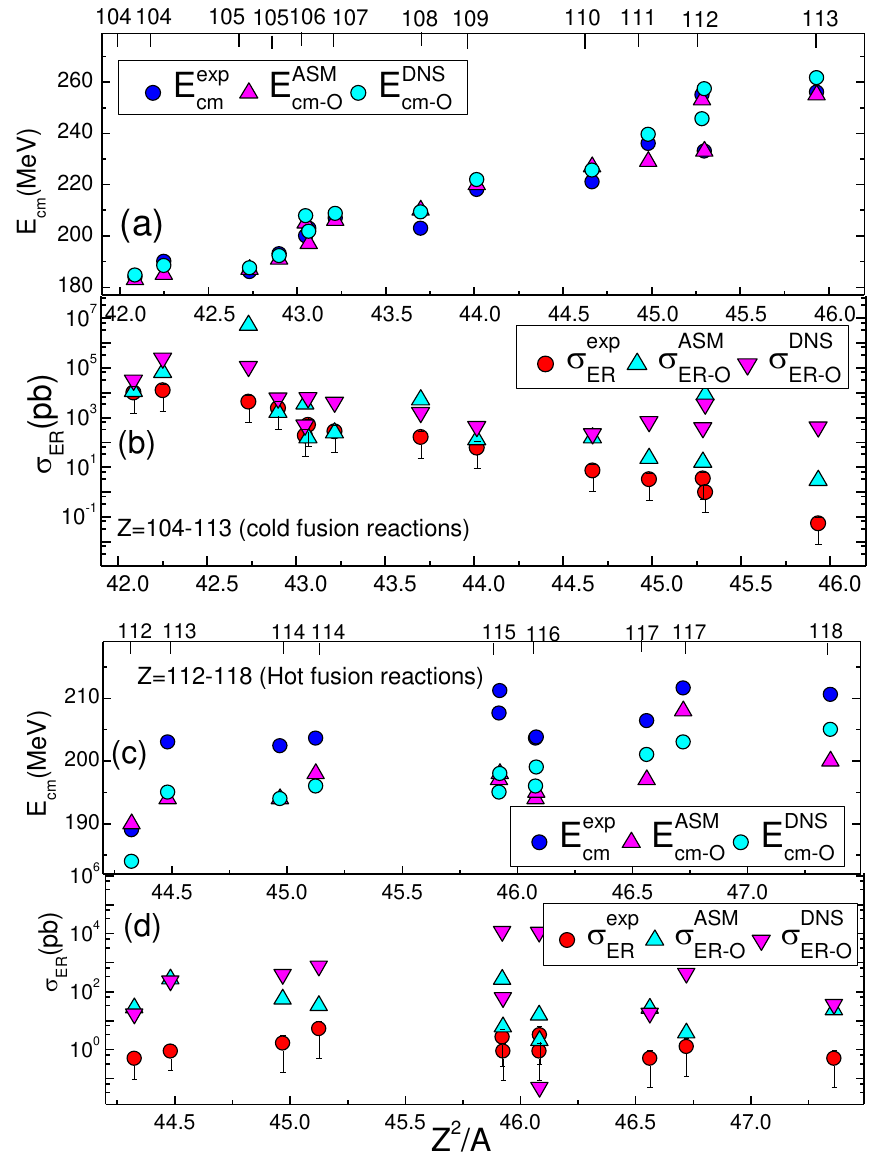}
\caption{Top panel: Optimal beam energies (a) and corresponding $\sigma_{ER}$ (b) as predicted by the ASM and DNS models for the cold  fusion reactions used to synthesize the superheavy elements $Z=104$ to 113  versus $Z^2/A$ plot. $E_{cm}^{exp.}$ indicates the beam energy used for the corresponding experiment. $\sigma_{ER-E}^{ASM}$ and $\sigma_{ER-E}^{DNS}$ are the expected ER cross section for different neutron evaporation channels at that energy.  Whereas $E_{cm-o}^{ASM}$ and $E_{cm-o}^{DNS}$ represent the optimal energies in centre of mass and $\sigma_{ER-O}^{ASM}$ and $\sigma_{ER-O}^{DNS}$ are the expected evaporation residue cross sections at the corresponding optimal energies predicted by the ASM and DNS models, respectively. Bottom panel: (c) and (d) are same as the panels (a) and (b), respectively, but here the hot fusion reactions are used to synthesize the superheavy elements $Z=112$ to 118.}
    \label{Cold and hot fusionreactions}
\end{figure}
\indent It is evident from equations \ref{sigmaEr-ASM} and \Ref{sigmaEr-DNS} that the  $\sigma_{ER}$ varies with $E_{cm}$ irrespective of application of the ASM or DNS model. Hence, these models can be used for calculating the excitation function for any heavy-ion fusion reactions. We have done so for the cold and hot fusion reactions which have been used to synthesize the SHEs $Z=104-118$. We have obtained the excitation function curves for various reactions meant for a specific SHE (a particular isotope) throughout the range $Z=104-118$. Four representative cases viz., for $Z=105,109,113,117$, are shown in Fig. \ref{Exc-func-ASM}. Here the calculation has been performed using the ASM. Similarly, the same calculations done by the DNS model and the corresponding plot is shown in supplemental material \cite{SM}. Such excitation function curves give clear indications about the most preferred choice or the optimal beam energy for a particular reaction. The optimal energy and the corresponding $\sigma_{ER}$ obtained from the ASM and DNS models are plotted as a function of $Z^2/A$ in Fig.\ref{Cold and hot fusionreactions}, $Z$ and $A$ are atomic number and mass number respectively for the compound SHN. The quantity $Z^2/A$ is chosen to accommodate different reactions used to synthesize the different isotopes of the same SHE. The beam energy used in the experiment and the respective $\sigma_{ER}$ are also included in Fig. \ref{Cold and hot fusionreactions} to compare with the model predictions. These plots imply that the beam energies used for the experiments were not in accordance with the ASM or DNS model predicted energies. However, for some cases the experimental beam energies were closed to the model predicted optimal energies. Interestingly, for those occasions the agreement between the measured and optimal $\sigma_{ER}$ is pretty good and thus, validity of the model predictions is established. Note that the data shown in Fig. \ref{Cold and hot fusionreactions} are also given in Table I of the Supplemental Material \cite{SM}.\\
\begin{figure}
    \centering
    \includegraphics[width=\linewidth]{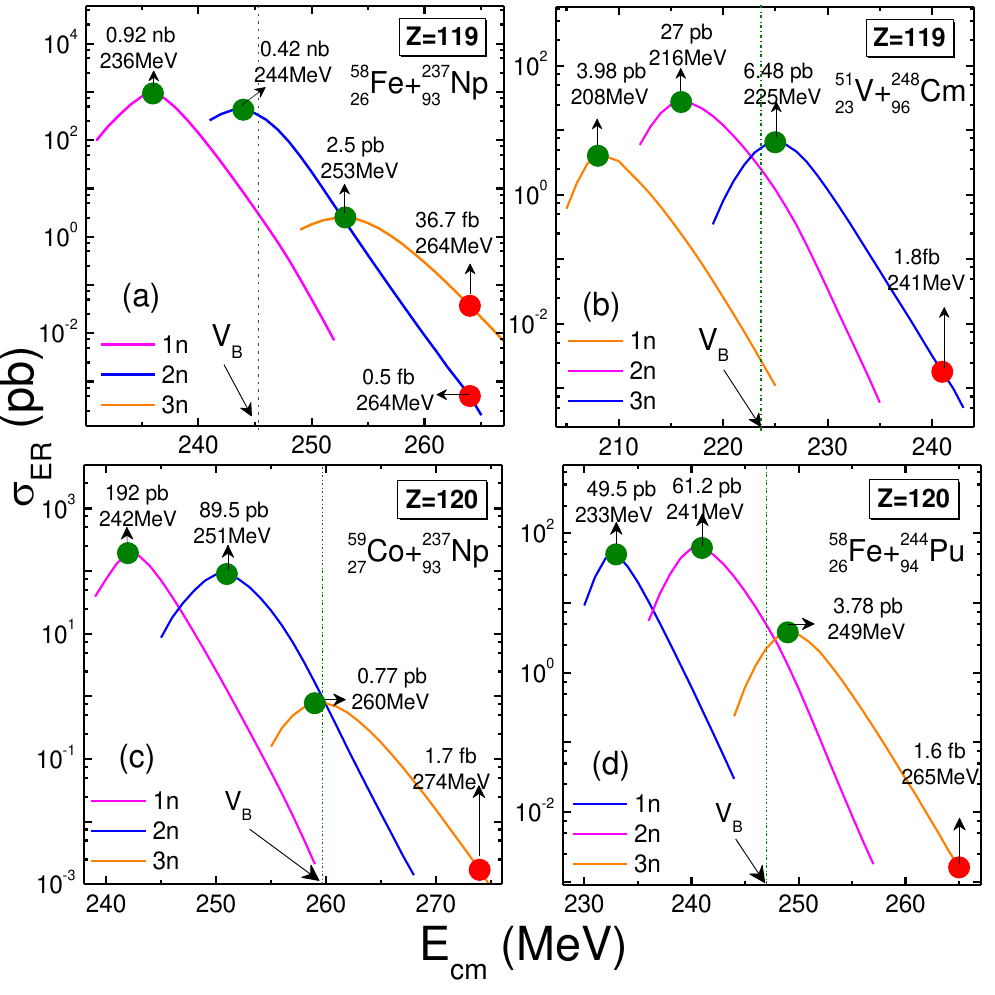}
    \caption{Excitation function curves for four superheavy reactions used to attempt synthesis of elements $Z= 119$ and 120, where the $\sigma_{ER}$ are obtained using the ASM. Other descriptions follow Fig. \ref{Exc-func-ASM}}.
    \label{failure119120-4panel}
\end{figure}
\indent In next stage, we intend to use above models in synthesizing SHE $Z>118$. Before doing so we would like to validate our DNS model calculations by reproducing an earlier known results. Nasirov $et. al.$ \cite{nasirov2009quasifission} used the DNS model to examine the reactions $^{54}Cr + ^{248}Cm$, $^{58}Fe + ^{244}Pu$ and $^{64}Ni + ^{238}U$ and found the first reaction is more favorable than the other two in synthesizing the SHE $Z=120$. Here, the capture stage was handled by the dynamical model and the fusion stage was tackled by the advanced statistical approach. This result was confirmed recently by an experiment \cite{novikov2020investigation} too. Further, another application of the said DNS model \cite{nasirov2011effects} inferred that the reaction $^{50}Ti+^{249}Cf$ is more favorable than the reaction $^{54}Cr + ^{248}Cm$. Here we verify this result with our DNS calculations and a comparative study is given in Table I. Very good agreement on the overall trend can be noticed between the earlier calculation \cite{nasirov2011effects} and the present evaluation. However, the certain difference of numbers between the two works 
are attributed to the following facts. \textcite{nasirov2011effects} used nuclear mass table from \textcite{moller1994stability} and \textcite{moller1994stability}. Whereas we used \textcite{wang2017ame2016} and  \textcite{moller2009heavy}, respectively, for the mass table and fission barrier.\\ 
\begin{figure}
  \centering
 \includegraphics[width=\linewidth]{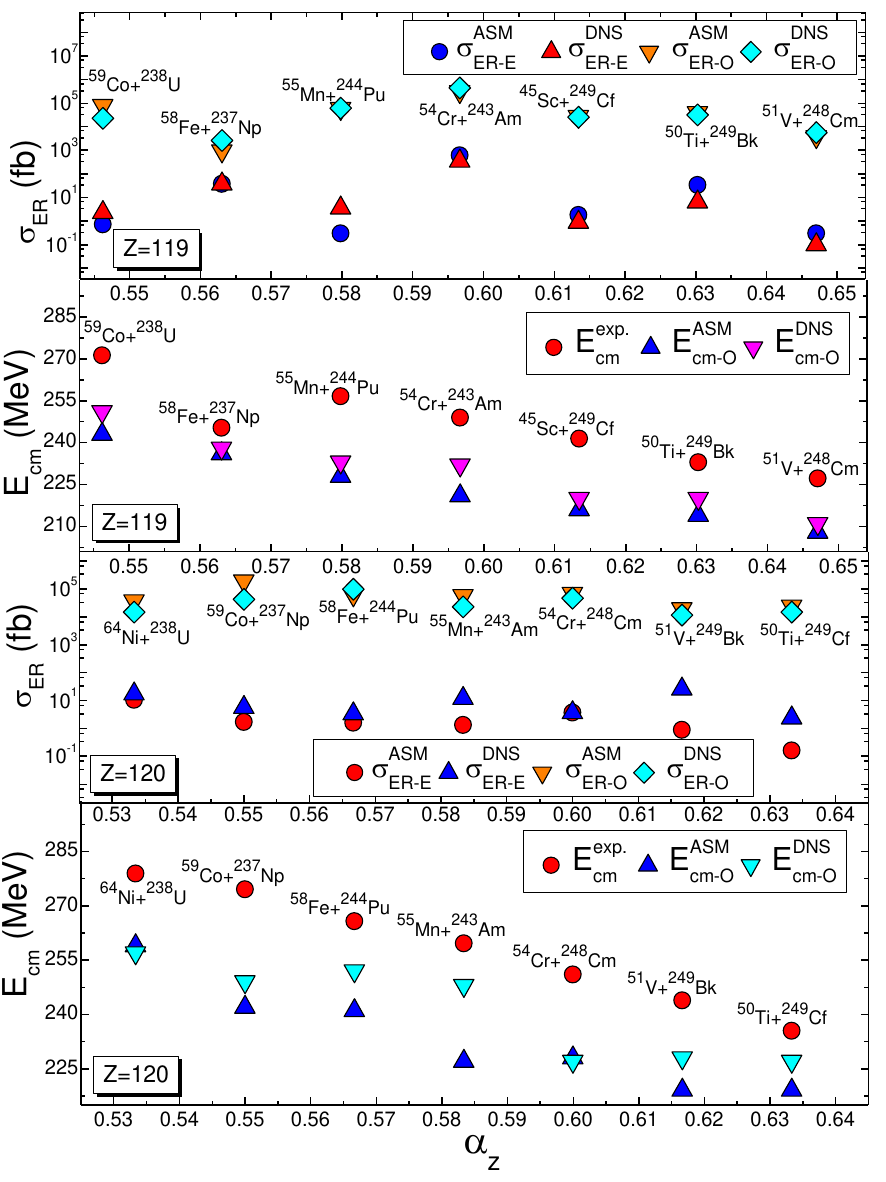}
 \caption{Optimal beam energies and corresponding $\sigma_{ER}$ as predicted by the ASM and DNS models for the massive fusion reactions used to synthesize the SHEs $Z=119$ and 120 are plotted as a function of charge asymmetry parameter $\alpha_Z$. Beam energies used for the experiments and corresponding $\sigma_{ER}$ calculated by the ASM and DNS models are also included. Symbols carry same meaning as specified in Fig.\ref{Cold and hot fusionreactions}.}
 \label{Z=119 and 120 ASm and DNS}
\end{figure}
\begin{table}
\caption{Comparison of the 3n ($\sigma_{ER}^{3n}$) and 4n ($\sigma_{ER}^{4n}$) evaporation residue cross sections in $fb$ units at different beam energies in centre of mass ($E_{cm}$) for the reactions $^{50}Ti+^{249}Cf$ and $^{54}Cr+^{248}Cm$ \cite{nasirov2011effects} with the present DNS model calculations.} 
\begin{tabular}{|c|c|c|c|c|c|c|}
\hline
\multirow{2}{*}{Reaction} & \multirow{2}{*}{$E_{cm}$} & \multicolumn{2}{c|}{$\sigma^{3n}_{ER}$} & \multirow{2}{*}{$E_{cm}$} & \multicolumn{2}{c|}{$\sigma^{4n}_{ER}$} \\ \cline{3-4} \cline{6-7} 
& & Prev{\cite{nasirov2011effects}} & Present     & & Prev{\cite{nasirov2011effects}} & Present     \\ \hline
\multirow{4}{*}{$^{50}_{22}Ti+^{249}_{98}Cf$} & 225 & 100 & 38.6 & 231.5 & 2.5 & 0.9 \\ \cline{2-7} 
 & 227.5 & 760 & 582 & 232.5 & 49 & 40 \\ \cline{2-7} 
 & 231.5 & 60 & 33.2 & 239.0 & 9.1 & 28 \\ \cline{2-7} 
 & 236.0 & 1.5 & 1.6 & 241.0 & 40 & 28.4 \\ \hline
\multirow{4}{*}{$^{54}_{24}Cr+^{248}_{96}Cm$} & 237.2 & 55 & 61.9 & 241.0 & 13 & 34.2 \\ \cline{2-7}
 & 241.5 & 76 & 118 & 249.6 & 28 & 14.4 \\ \cline{2-7} 
 & 246.7 & 14 & 17.6 & 252.0 & 12 & 3.5 \\ \cline{2-7} 
 & 248.2 & 0.2 & 0.1 & - & - & - \\ \hline

\end{tabular}
\label{nasirov}
\end{table}
\begin{figure}
    \centering
    \includegraphics[width=\linewidth]{{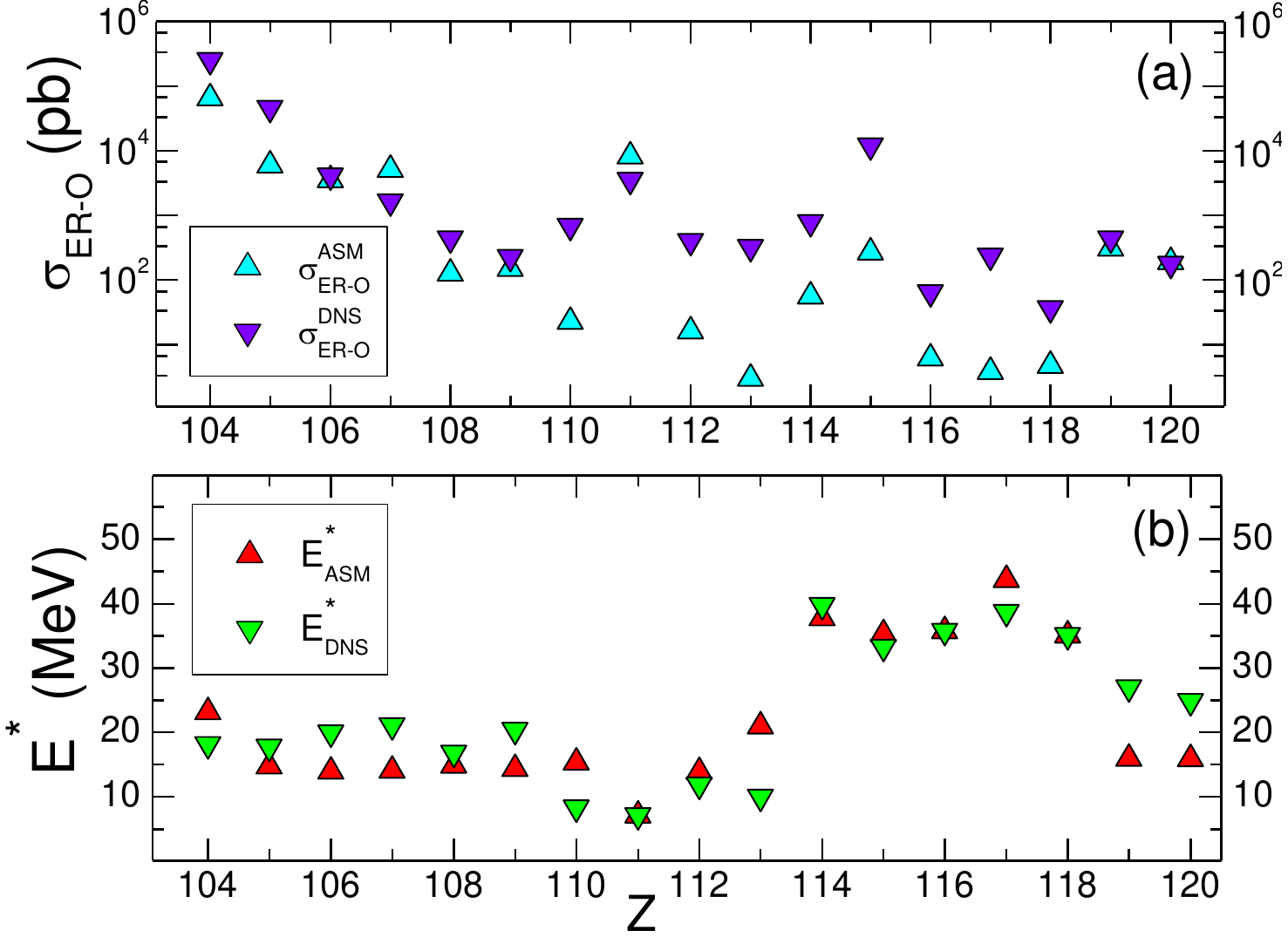}}
    \caption{Optimal $\sigma_{ER}$ ($\sigma_{ER-O}$) predicted by the ASM and DNS models (a) and corresponding excitation energy ($E^*$) as a function of atomic number of the SHE for a particular superheavy reaction as marked in Table I and II of the Supplemental Material \cite{SM}. }
    \label{Zvsereavg2stack}
\end{figure}
\indent We have evaluated the optimal energies and corresponding $\sigma_{ER}$ using the ASM and DNS models for all the reactions used to attempt for synthesizing SHE $Z>118$ and we show a few such cases in Fig.\ref{failure119120-4panel}. The predicted cross section at every beam energy used for the experiment is also marked. It reveals a point that the beam energies used were much higher than the corresponding predicted optimal energies and thus, it resulted too low $\sigma_{ER}$ to measure in the experiments. Further, we scrutinize here the fate of all these experiments if were done at the optimal energies. We choose one of the entrance channel parameters to plot the $\sigma_{ER}$ values at experimental and optimal beam energies. Here it is the charge asymmetry parameter $\alpha_z=\lvert(Z_1-Z_2)/(Z_1+Z_2)\rvert$ \cite{manjunatha2020entrance}. The  optimal $\sigma_{ER}$ are plotted against $\alpha_z$ in Fig. \ref{Z=119 and 120 ASm and DNS}. Further, the corresponding data are also listed in Table II of Supplemental Material \cite{SM}. It can clearly be seen that only a few out of 14 reactions lead to $<$ 1 $pb$ ER cross section and quite a few reactions even yield $>$ 100 $pb$. According to the ASM as well as DNS models the best reaction for SHE $Z=119$ is $^{54}Cr+^{243}Am\rightarrow^{297}119$ and that for SHE $Z=120$ is $^{59}Co+^{237}Np\rightarrow^{296}120$.\\
\indent Till date many superheavy reactions have been used even for synthesizing a single SHE but its different isotopes. Let us consider to synthesize the SHE first, not any particular isotope, which gives us maximum production cross section. In this view point, in first step, we study the optimal $\sigma_{ER}$ ($\sigma_{ER-O}$) for all the reactions used for the synthesis of $Z=104-120$ using the ASM and DNS models as shown in Figs. \ref{Cold and hot fusionreactions} and \ref{failure119120-4panel} (also in Table I and II of \cite{SM}, respectively). In second step, we pick up the reaction for a specific SHE that gives us the largest  $\sigma_{ER-O}$ and plot the corresponding $\sigma_{ER-O}$ as a function of Z of the compound nucleus as shown in Fig. \ref{Zvsereavg2stack}. We provide the corresponding excitation energies ($E^*$) in this figure also. We can notice from this figure that the trend of the $\sigma_{ER}$ and $E^*$ predicted by the ASM and DNS models are similar. The DNS model predicts higher production cross section than that of the ASM in most of the cases. Whereas no such order for the prediction of $E^*$ can be noticed. \\
\indent One fact is well known that the entrance channel plays an important role in choosing a superheavy reaction and thus a priory to conducting the experiment we must first judge on the reaction. It can now be done well with one of our recent works \cite{Rules}. According to it, all the superheavy reactions employed for the synthesis of SHEs $Z=119$ and 120 pass these criteria. Despite, all attempts were failed \cite{dullmann2016search,oganessian2009attempt,hofmann2015super}! Now, the present work reveal a very significant fact  that the beam energies used were much higher than the predicted optimal energies for synthesizing SHEs $Z=119$ and 120 as shown in Fig. \Ref{Z=119 and 120 ASm and DNS} and also in Table II of \cite{SM}. It is evident from Fig. \ref{Zvsereavg2stack} that the expected $\sigma_{ER-O}$ and  $E^*$ for SHEs $Z=119$ and 120 are higher and lower, respectively than that for neighboring SHEs $Z=116-118$. However, higher  $E^*$ was used in the experiments for the synthesis of $Z=119$ and 120. Origin of such action may be due to the fact that excitation function curves of heavy ion reactions show a saturation or slowly decreasing trend on $\sigma_{ER}$ with $E_{cm}$ at the region $E_{cm} > V _B$ \cite{reisdorf1985fusability,brinkmann1994residue,hinde2002severe,mukherjee2007failure,gehlot2019evaporation,laveen2020fusion}. In contrast the excitation function curves of superheavy ion reactions exhibit a bell shaped nature as shown in Fig. \ref{Exc-func-ASM} and \ref{failure119120-4panel}. A recent \cite{tanaka2020study} and many earlier experiments \cite{hofmann1997excitation,hessberger2001decay,oganessian2004measurements,ts2006synthesis} corroborate this fact well. A thorough comparison of ASM and DNS excitation functions as a function of Excitation energy with several experiments are shown in Fig.4 of the Supplemental Material \cite{SM}. Thus, higher energy used was the reason of the setback.   \\
\indent To conclude, we have employed the ASM and DNS models to study the excitation function of number of superheavy reactions used for synthesizing the SHEs $Z=104-120$. Both the models support each other and this fact helps us to infer that proper choice of optimal excitation energy for every superheavy reaction takes a supreme role to yield the largest production cross section. Exploitation of this conclusion may save the accelerator time to a great extent and thus emancipate the financial loads. On the other hand, might enhance greatly the measurement statistics. Very significantly, this study reveals that improper choice of beam energy was the main cause of the debacle of all the experiments attempted for synthesis of SHEs $Z=119-120$.  Now, if experiments are conducted at the proposed optimal energies, the same superheavy reactions might lead to synthesizing the SHEs $Z=119$ and 120 with optimal $\sigma_{ER}$ more than a hundred of $pb$.   
\bibliographystyle{apsrev4-2}
\bibliography{apssamp}
\end{document}